\documentclass{article}

     \usepackage[nonatbib,preprint]{neurips_2020}

\usepackage[utf8]{inputenc} 
\usepackage[T1]{fontenc}    
\usepackage{hyperref}       
\usepackage{url}            
\usepackage{booktabs}       
\usepackage{amsfonts}       
\usepackage{nicefrac}       
\usepackage{microtype}      
\usepackage{graphicx}       
\usepackage{subcaption}     

\title{COVID-Net~CT-2: Enhanced Deep Neural Networks for Detection of COVID-19 from Chest CT Images Through Bigger, More Diverse Learning}

\author{
    Hayden Gunraj$^{1,*}$, Ali Sabri$^{5}$, David Koff$^{6}$,  and Alexander Wong$^{2,3,4,*}$\\
    $^{1}$Department of Mechanical and Mechatronics Engineering, University of Waterloo, Canada\\
    $^{2}$Department of Systems Design Engineering, University of Waterloo, Canada\\
    $^{3}$Waterloo Artificial Intelligence Institute, Canada\\
    $^{4}$DarwinAI Corp., Canada\\
    $^{5}$Department of Radiology, Niagara Health, McMaster University, Canada\\
    $^{6}$Department of Radiology, Hamilton Health Sciences, McMaster University, Canada\\
    $^{*}$Corresponding authors: \texttt{\{hayden.gunraj,a28wong\}@uwaterloo.ca}\\
}

\begin{document}

\maketitle

\begin{abstract}
    \textbf{Background:} The COVID-19 pandemic continues to rage on around the world, with multiple waves causing substantial harm to health and economies around the world.  Motivated by the use of computed tomography (CT) imaging at clinical institutes around the world as an effective complementary screening method to RT-PCR testing, we introduced COVID-Net~CT, a deep neural network tailored for detection of COVID-19 cases from chest CT images, along with a large curated benchmark dataset comprising 1,489 patient cases as part of the open source COVID-Net initiative. However, one potential limiting factor is restricted quantity and diversity given the single nation patient cohort used in the study.\\
    \textbf{Methods}: Motivated by the success of COVID-Net~CT, we introduce COVID-Net~CT-2, enhanced deep neural networks for COVID-19 detection from chest CT images trained on the largest quantity and diversity of multinational patient cases in research literature.  We accomplish this through the introduction of two new CT benchmark datasets, the largest of which comprises a multinational cohort of 4,501 patients from at least 15 countries. To the best of our knowledge, this represents the largest, most diverse multinational cohort for COVID-19 CT images in open access form. We leverage explainability to investigate the decision-making behaviour of COVID-Net~CT-2 to ensure that decisions are based on relevant indicators, with the results for select cases reviewed and reported on by two board-certified radiologists with over 10 and 30 years of experience, respectively.\\
    \textbf{Results:} The COVID-Net~CT-2 neural networks achieved accuracy, COVID-19 sensitivity, positive predictive value, specificity, and negative predictive value of 98.1\%/96.2\%/96.7\%/99\%/98.8\% and 97.9\%/95.7\%/96.4\%/98.9\%/98.7\%, respectively. Explainability-driven performance validation shows that COVID-Net~CT-2's decision-making behaviour is consistent with radiologist interpretation by leveraging correct, clinically relevant critical factors.\\
    \textbf{Conclusions:}
    The results are promising and suggest the strong potential of deep neural networks as an effective tool for computer-aided COVID-19 assessment. While not a production-ready solution, we hope the open-source, open-access release of COVID-Net~CT-2 and benchmark datasets\footnote{\url{https://github.com/haydengunraj/COVIDNet-CT}} will continue to enable researchers, clinicians, and citizen data scientists alike to build upon them.

\end{abstract}
\section{Introduction}
    The coronavirus disease 2019 (COVID-19) pandemic, caused by severe acute respiratory syndrome coronavirus 2 (SARS-CoV-2), continues to rage on around the world, with multiple waves causing substantial harm to health and economies around the world. Real-time reverse transcription polymerase chain reaction (RT-PCR) testing remains the primary screening tool for COVID-19, where SARS-CoV-2 ribonucleic acid (RNA) is detected within an upper respiratory tract sputum sample~\cite{Wang2020_RTPCR}.  However, despite being highly specific, the sensitivity of RT-PCR can be relatively low~\cite{Fang2020, Li2020_RTPCR} and can vary greatly depending on the time since symptom onset as well as sampling method~\cite{Yang2020, Li2020_RTPCR, Ai2020}.

    Clinical institutes around the world have explored the use of computed tomography (CT) imaging as an effective, complementary screening tool alongside RT-PCR~\cite{Fang2020, Ai2020, Xie2020}.  In particular, studies have shown CT to have great utility in detecting COVID-19 infections during routine CT examinations for non-COVID-19 related reasons such as elective surgical procedure monitoring and neurological examinations~\cite{Tian,Shatri}.  Other scenarios where CT imaging has been leveraged include cases where patients have worsening respiratory complications, as well as cases where patients with negative RT-PCR test results are suspected to be COVID-19 positive due to other factors.  Early studies have shown that a number of potential indicators for COVID-19 infections may be present in chest CT images~\cite{Guan2020, Wang2020, Chung2020, Pan2020, Fang2020, Ai2020, Xie2020}, but may also be present in non-COVID-19 infections. This can lead to challenges for radiologists in distinguishing COVID-19 infections from non-COVID-19 infections using chest CT~\cite{Bai2020_Perf, Mei2020}.

    Inspired by the potential of CT imaging as a complementary screening method and the challenges of CT interpretation for COVID-19 screening, we previously introduced COVID-Net~CT~\cite{Gunraj2020}, a deep convolutional neural network  tailored for detection of COVID-19 cases from chest CT images.  We further introduced COVIDx~CT, a large curated benchmark dataset comprising of chest CT scans from a cohort of 1,489 patients derived from a collection by the China National Center for Bioinformation (CNCB)~\cite{cncb}.  Both COVID-Net~CT and COVIDx~CT were made publicly available as part of the COVID-Net~\cite{covidnet,alex2020covidnets} initiative, an open source initiative\footnote{\url{https://alexswong.github.io/COVID-Net}} aimed at accelerating advancement and adoption of deep learning in the fight against the COVID-19 pandemic.  While COVID-Net~CT was able to achieve state-of-the-art COVID-19 detection performance, one potential limiting factor is the restricted quantity and diversity of CT imaging data used to learn the deep neural network given the entirely Chinese patient cohort used in the study.  As such, a greater quantity and diversity in the patient cohort has the potential to improve generalization, particularly when COVID-Net~CT is leveraged under different clinical settings around the world.

    Motivated by the success and widespread adoption of COVID-Net~CT and COVIDx~CT, as well as its potential data quantity and diversity limitations, in this study we introduce COVID-Net~CT-2, enhanced deep convolutional neural networks for COVID-19 detection from chest CT images that are trained on a large, diverse, multinational patient cohort.  More specifically, we accomplish this through the introduction of two new CT benchmark datasets (COVIDx~CT-2A and COVIDx~CT-2B), the largest of which comprises a multinational cohort of 4,501 patients from at least 15 countries. To the best of the authors' knowledge, these benchmark datasets represent the largest, most diverse multinational cohorts for COVID-19 CT images available in open access form. Finally, we leverage explainability to investigate the decision-making behaviour of COVID-Net~CT-2 to ensure decisions are based on relevant visual indicators in CT images, with the results for select patient cases being reviewed and reported on by two board-certified radiologists with 10 and 30 years of experience, respectively. The COVID-Net~CT-2 networks and corresponding COVIDx~CT-2 datasets are publicly available as part of the COVID-Net initiative~\cite{covidnet,alex2020covidnets}. While not a production-ready solution, we hope the open-source, open-access release of the COVID-Net~CT-2 networks and the corresponding COVIDx~CT-2 benchmark datasets will enable researchers, clinicians, and citizen data scientists alike to build upon them.

    \begin{figure}[t]
        \centering
        \includegraphics[width=\textwidth]{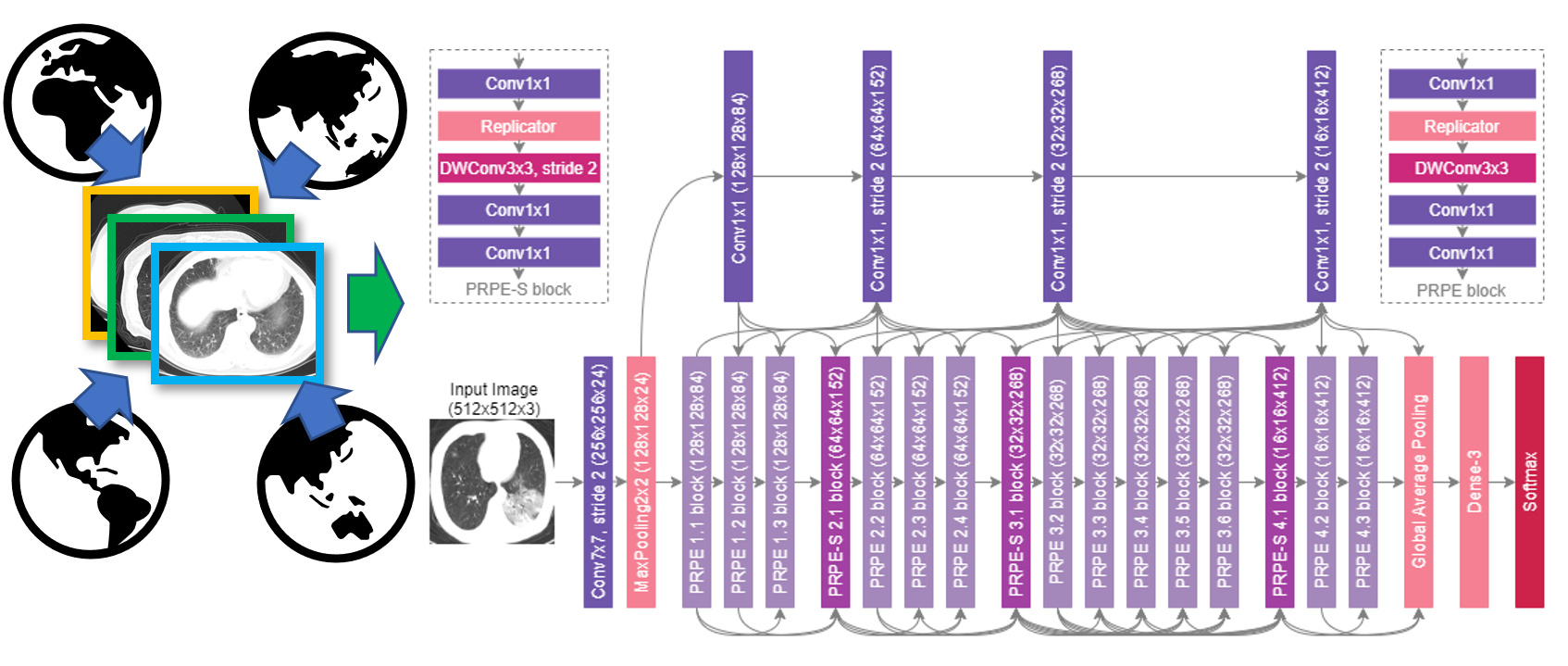}
        \caption{COVID-Net~CT-2 architecture design and COVIDx~CT-2 benchmark.  We leverage the COVID-Net~CT network architecture~\cite{Gunraj2020} as the core of the COVID-Net~CT-2 networks (COVID-Net~CT-2~L network shown in figure, with COVID-Net~CT-2~S network sharing same macroarchitecture design but fewer parameters), which was discovered automatically via machine-driven design exploration. Some interesting characteristics about the COVID-Net~CT-2 design include a very diverse yet lightweight designs ($\sim$16.8$\times$ and $\sim$52.6$\times$ lower architectural complexity than ResNet-50~\cite{resnet} for COVID-Net~CT-2~L and S networks, respectively) and selective long-range connectivity to draw a balance between complexity and representational power.  The COVID-Net~CT-2 design was trained on CT scans from a large, diverse, multinational cohort of patient cases across at least 15 countries (i.e., COVIDx~CT-2).}
        \label{fig:arch}
    \end{figure}
\section{Methods}\label{methods}
    In this study, we introduce COVID-Net~CT-2~L and COVID-Net~CT-2~S, a pair of enhanced deep convolutional neural networks for the detection of COVID-19 from chest CT. To train and test these networks, we further introduce two COVIDx~CT-2 benchmark datasets which represent the largest, most diverse multinational patient cohorts for COVID-19 CT images available in open access form, spanning cases from at least 15 countries. A visual overview of COVID-Net~CT-2 and COVIDx~CT-2 is shown in Figure~\ref{fig:arch}. The methodology behind the preparation of the two COVIDx~CT-2 benchmark datasets, the construction and learning of the COVID-Net~CT-2 networks, and the explainability-driven performance validation process are described in detail below.

\subsection{COVIDx~CT-2 benchmark dataset preparation}

    The original COVIDx~CT benchmark dataset consists of chest CT scans collected by the China National Center for Bioinformation (CNCB)~\cite{cncb} which were carefully processed and selected to form a cohort of 1,489 patient cases.  While COVIDx~CT is significantly larger than many CT datasets for COVID-19 detection in literature, a potential limitation with leveraging COVIDx~CT for learning deep neural networks is the limited diversity in terms of patient demographics.  More specifically, the cohort of patients used in COVIDx~CT are collected in different provinces of China, and as such the characteristics of COVID-19 infection as observed in the chest CT images may not generalize to patients around the world outside of China. Therefore, increasing the quantity and diversity of the patient cohort in constructing new benchmark datasets could result in more diverse, well-rounded learning of deep neural networks. In doing so, improved generalization and applicability for use under different clinical environments around the world can be achieved.

    In this study, we carefully processed and curated CT images from several patient cohorts from around the world which were collected using a variety of CT equipment types, protocols, and levels of validation. By unifying CT imaging data from several cohorts from around the world, we created two diverse, large-scale benchmark datasets:
    \begin{itemize}
        \item \textbf{COVIDx~CT-2A}: This benchmark dataset comprises 194,922 CT images from a multinational cohort of 3,745 patients between 0 and 93 years old (median age of 51) with strongly clinically-verified findings. The multinational cohort consists of patient cases collected by the following organizations and initiatives from around the world: (1) China National Center for Bioinformation (CNCB)~\cite{cncb} (China), (2) National Institutes of Health Intramural Targeted Anti-COVID-19 (ITAC) Program (hosted by TCIA~\cite{TCIA}, countries unknown), (3) Negin Radiology Medical Center~\cite{rahimzadeh2020fully} (Iran), (4) Union Hospital and Liyuan Hospital of Huazhong University of Science and Technology~\cite{HUST} (China), (5) COVID-19 CT Lung and Infection Segmentation initiative, annotated and verified by Nanjing Drum Tower Hospital~\cite{COVID-19-SegBenchmark} (Iran, Italy, Turkey, Ukraine, Belgium, some countries unknown), (6) Lung Image Database Consortium (LIDC) and Image Database Resource Initiative (IDRI)~\cite{LIDC} (countries unknown), and (7) Radiopaedia collection~\cite{radiopaedia} (Iran, Italy,  Australia, Afghanistan, Scotland, Lebanon, England, Algeria, Peru, Azerbaijan, some countries unknown).
        \item \textbf{COVIDx~CT-2B}: This benchmark dataset comprises 201,103 CT images from a multinational cohort of 4,501 patients between 0 and 93 years old (median age of 51) with a mix of strongly verified findings and weakly verified findings. The patient cohort in COVIDx~CT-2B consists of the multinational patient cohort we leveraged to construct COVIDx~CT-2A, which have strongly clinically-verified findings, with additional patient cases with weakly verified findings collected by the Research and Practical Clinical Center of Diagnostics and Telemedicine Technologies, Department of Health Care of Moscow (MosMed)~\cite{Morozov2020.05.20.20100362} (Russia). Notably, these additional cases are only included in the training dataset, and as such the validation and test datasets are identical to those of COVIDx~CT-2A.
    \end{itemize}

\begin{figure}[t]
        \centering
        \includegraphics[width=\textwidth]{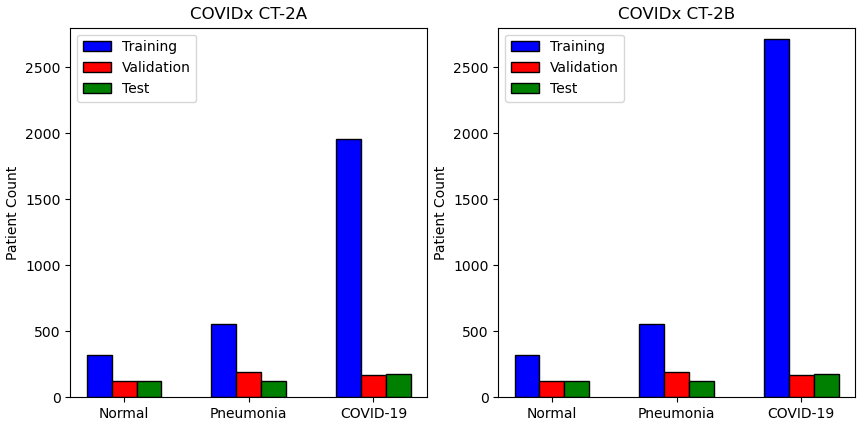}
        \caption{Patient distribution across training, validation, and test for COVIDx~CT-2A (left) and COVIDx~CT-2B (right).}
        \label{fig:distribution}
    \end{figure}

    In both COVIDx~CT-2 benchmark datasets, the findings for the chest CT volumes corresponds to three different infection types: (1) novel coronavirus pneumonia due to SARS-CoV-2 viral infection (NCP), (2) common pneumonia (CP), and (3) normal controls, with the patient distribution for the three infection types across training, validation, and test shown in Figure~\ref{fig:distribution}. For CT volumes labelled as NCP or CP, slices containing abnormalities were identified and assigned the same labels as the CT volumes. Notably, patient age was not available for all cases, and as such the age ranges and median ages reported above are based on patient cases for which age was available. For images which were originally in Hounsfield units (HU), a standard lung window centered at -600~HU with a width of 1500~HU was used to map the images to unsigned 8-bit integer range (i.e., [0, 255]).

\begin{figure}[t]
        \centering
        \includegraphics[width=\textwidth]{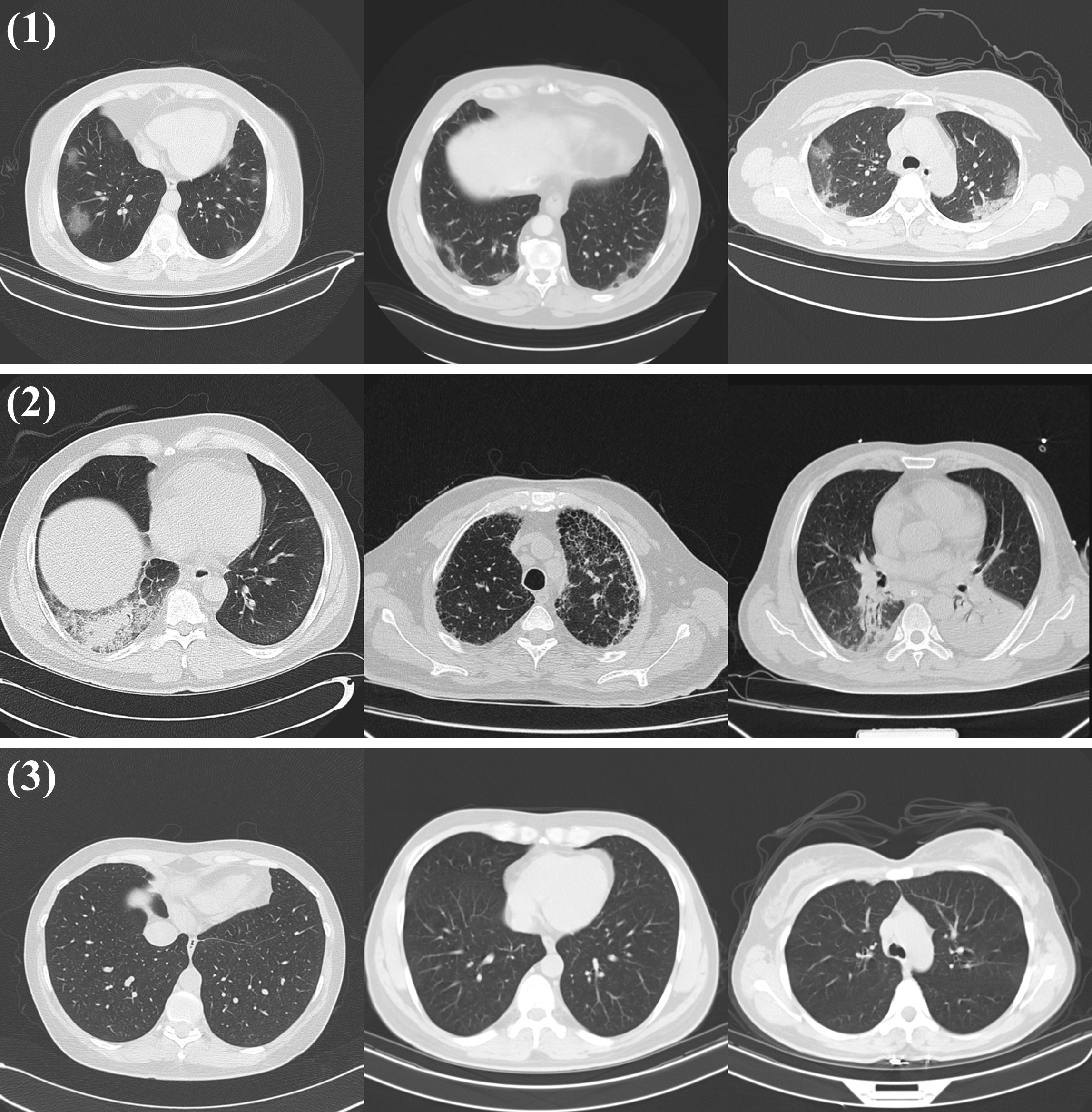}
        \caption{Example CT images from the COVIDx~CT-2 benchmark datasets from each type of infection: (1) normal coronavirus pneumonia due to SARS-CoV-2 infection (NCP), (2) common pneumonia (CP), and (3) normal controls.}
        \label{fig:examples}
    \end{figure}

    The rationale for creating two different COVIDx~CT-2 benchmark datasets stems from the availability of weakly verified findings (i.e., findings not based on RT-PCR test results or final radiology reports), which can be useful for further increasing the quantity and diversity of patient cases that a deep neural network can be exposed to and can be of great interest for researchers, clinicians, and citizen scientists to explore and build upon while being made aware of the fact some of the CT scans do not have strongly verified findings available. Both COVIDx~CT-2A and COVIDx~CT-2B benchmark datasets are publicly available\footnote{\url{https://www.kaggle.com/hgunraj/covidxct}} as part of the COVID-Net initiative, with example CT images from each type of infection shown in Figure~\ref{fig:examples}.

\subsection{COVID-Net~CT-2 construction and learning}

    By leveraging the COVIDx~CT-2 benchmark datasets introduced in the previous section, we build the COVID-Net~CT-2 deep convolutional neural networks in a way that is more generalizable and more readily adoptable to a wider range of clinical scenarios around the world through bigger, more diverse learning on the largest quantity and diversity of multinational patient cases in research literature.  More specifically, two COVID-Net~CT-2 networks are built (COVID-Net~CT-2~L and COVID-Net~CT-2~S), with both sharing the same macroarchitecture design but different number of parameters.  The COVID-Net~CT-2 architecture is shown in Figure~\ref{fig:arch}, and the networks are made publicly available\footnote{\url{https://github.com/haydengunraj/COVIDNet-CT}}. More specifically, we leverage the COVID-Net~CT network architecture design proposed in~\cite{Gunraj2020} as the core of the architecture designs of the COVID-Net~CT-2 networks. The architecture designs were discovered automatically via a machine-driven design exploration process using generative synthesis~\cite{gensynth}, where the macroarchitecture and microarchitecture designs of a tailored deep neural network architecture for the task and data at hand is determined via iterative constrained optimization based on a universal performance function (e.g.,~\cite{wong2019netscore}) and a set of quantitative constraints. The result is highly customized architecture designs that strike a strong balance between complexity and representational power beyond what a human designer can achieve alone.

    The COVID-Net~CT-2 designs possess several interesting architectural characteristics.  First, COVID-Net~CT-2 designs exhibit very diverse yet lightweight designs composed largely of a heterogeneous combination of strided and unstrided depthwise convolutions as well as pointwise convolutions, with unique microarchitecture design characteristics tailored during the machine-driven design exploration process.  Second, COVID-Net~CT-2 leverages selective long-range connectivity through several point convolution hubs to draw a balance between architectural complexity and representational power.  As a result of these macroarchitecture and microarchitecture design traits tailored around COVID-19 detection from CT images, the COVID-Net~CT-2 architecture designs have, at $\sim$1.4M parameters and $\sim$0.45M parameters, approximately $\sim$16.8$\times$ and $\sim$52.6$\times$ lower architectural complexity than ResNet-50~\cite{resnet}, respectively.

    The constructed COVID-Net~CT-2 deep convolutional neural networks were trained on the COVIDx~CT-2A benchmark dataset via stochastic gradient descent with momentum~\cite{momentum}, where the following hyperparameters were leveraged: learning rate=5e-4, momentum=0.9, number of epochs=25, batch size=64. To further increase data diversity beyond what is provided by the large multinational cohort in order to improve the generalization of COVID-Net~CT-2, we leveraged data augmentation in the form of cropping box jitter, rotation,
    horizontal and vertical shear, horizontal flip, and intensity shift and scaling. The construction, training, and evaluation of COVID-Net~CT-2 networks were conducted using the TensorFlow~\cite{tensorflow} machine learning library.

\subsection{Explainability-driven performance validation via model auditing}\label{xai}

    As with COVID-Net~CT~\cite{Gunraj2020}, we utilize GSInquire~\cite{gsinquire} as the explainability method of choice to conduct explainability-driven performance validation, Using GSInquire, we audit the trained COVID-Net~CT-2 to better understanding and verify its decision-making behaviour when analyzing CT images to predict the condition of a patient.  This form of performance validation via model auditing is particularly important in a clinical context, as the decisions made about a patient's conditions can affect the health of patients via treatment and care decisions made using a model's predictions.  Therefore, examining the decision-making behaviour through model auditing is key to ensuring that the right visual indicators in the CT scans (in the case of COVID-19 infections, visual anomalies such as ground-glass opacities and bilateral abnormalities) are leveraged for making a prediction as opposed to irrelevant visual cues (e.g., synthetic padding, circular scan artifacts, patient table, etc.).  Furthermore, incorporating interpretability in the validation process also increases the level of trust that a clinician has in leveraging such models for clinical decision support by adding an extra degree of algorithmic transparency.

    To facilitate explainability-driven performance validation via model auditing, GSInquire provides an explanation of how a model makes a decision based on input data by identifying a set of critical factors within the input data that impact the decision-making process of the deep neural network in a quantitatively significant way. This is accomplished by probing the
    model with an input signal (in this case, a CT image) as the targeted stimulus signal and observing the reactionary response signals throughout the model, thus enabling quantitative insights to be derived through the inquisition process. These quantitative insights are then transformed and projected into the same space as the input signal to produce an interpretation (in this case, a set of critical factors in the CT image that quantitatively led to the prediction of the patient's condition).  These interpretations can be visualized spatially relative to the CT images for greater insights into the decision-level behaviour of COVID-Net~CT-2.  Compared to other explainability methods~\cite{kumar2017explaining,lundberg2017unified,ribeiro2016why,erion2020improving,kumar2017discovery}, this interesting nature of GSInquire in identifying quantitative impactful critical factors enables it to achieve explanations that better reflect the decision-making process of models when compared to other state-of-the-art explainability methods~\cite{gsinquire}. This makes it particularly suitable for quality assurance of models prior to clinical deployment to identify errors, biases, and anomalies that can lead to `right decisions for the wrong reasons'.  The results obtained during the explainability-driven performance validation via model auditing for select patient cases are further reviewed and reported on by two board-certified radiologists (A.S. and D.K.). The first radiologist (A.S.) has over 10 years of experience, while the second radiologist (D.K.) has over 30 years of radiology experience.

\section{Results}\label{results}

\subsection{Quantitative analysis}

    To explore the efficacy of the COVID-Net~CT-2 networks for COVID-19 detection from CT images, we conducted a quantitative evaluation of the trained deep neural networks using the COVIDx~CT-2 test dataset.  For comparison purposes, we also conduct a quantitative comparison with COVID-Net~CT~\cite{Gunraj2020} (referred from here on as COVID-Net~CT-1 for clarity), which was previously shown to achieve state-of-the-art performance when compared with state-of-the-art deep neural network architectures such as ResNet-50~\cite{resnetv2}, NASNet-A-Mobile~\cite{nasnet}, and EfficientNet-B0~\cite{efficientnet} for the task of COVID-19 detection from CT images.  The test accuracy of the COVID-Net~CT-2 networks and COVID-Net~CT-1 are shown in Table~\ref{table:comp}. It can be observed that COVID-Net~CT-2~L and COVID-Net~CT-2~S achieved strong test accuracies of 98.1\% and 97.9\%, respectively, on the COVIDx~CT-2 test dataset, while at the same time possessing low architectural complexity ($\sim$1.4M parameters and $\sim$0.45M parameters, respectively) and low computational complexity ($\sim$4.18 GFLOPs and $\sim$1.94 GFLOPs).  Compared to COVID-Net~CT-1, it can be observed that COVID-Net~CT-2~S and COVID-Net~CT-2~L achieved 3.4\% and 3.6\% higher accuracy, respectively.

    The sensitivity and positive predictive value (PPV) for each infection type on the COVIDx~CT-2 test dataset is shown in Table~\ref{table:sens} and Table~\ref{table:ppv}, respectively.  It can be observed that COVID-Net~CT-2~L and COVID-Net~CT-2~S were able to achieve both high COVID-19 sensitivity (96.2\% and 95.7\%, respectively) and high COVID-19 PPV (96.7\% and 96.4\%, respectively). Compared to COVID-Net~CT-1, it can be observed that COVID-Net~CT-2~S and COVID-Net~CT-2~L achieved 15.5\% and 16\% higher COVID-19 sensitivity, respectively. At the cost of significantly lower COVID-19 sensitivity, COVID-Net~CT-1 is able to achieve 0.9\% and 1.2\% higher COVID-19 PPV than COVID-Net~CT-2~L and COVID-Net~CT-2~S, respectively. From a clinical perspective, high sensitivity ensures few false negatives which would lead to missed patients with COVID-19 infections, whereas high PPV ensures few false positives which add an unnecessary burden on the healthcare system, which is already stressed due to the ongoing pandemic.

    The specificity and negative predictive value (NPV) for each infection type on the COVIDx~CT-2 test dataset is shown in Table~\ref{table:spec} and Table~\ref{table:npv}, respectively.  It can be observed that COVID-Net~CT-2~L and COVID-Net~CT-2~S were able to achieve both high COVID-19 specificity (99\% and 98.9\%, respectively) and high COVID-19 NPV (98.8\% and 98.7\%, respectively). Compared to COVID-Net~CT-1, it can be observed that COVID-Net~CT-2~S and COVID-Net~CT-2~L achieved 4.5\% and 4.6\% higher COVID-19 NPV, respectively. Notably, COVID-Net~CT-1 achieves 0.4\% and 0.5\% higher COVID-19 specificity than COVID-Net~CT-2~L and COVID-Net~CT-2~S, respectively, but as previously mentioned this comes at the cost of significantly lower COVID-19 sensitivity. The high specificity and NPV achieved by COVID-Net~CT-2 are important from a clinical perspective to ensure that COVID-19-negative predictions are indeed true negatives in the vast majority of cases, which facilitates rapid identification of COVID-19-negative patients.

    These experimental results are particularly promising in terms of model generalization and applicability for use under different clinical environments given the much more diverse nature of the COVIDx~CT-2 test dataset. As such, these results demonstrate the potential value of COVID-Net~CT-2 as an effective clinical decision support tool to aid with COVID-19 screening.

    \begin{table}[t]
        \caption{Accuracy (image level) for the tested networks on the COVIDx~CT-2 benchmark test dataset. Best results highlighted in \textbf{bold}.}
        \medskip
        \label{table:comp}
        \centering
        \begin{tabular}{ll}
            \toprule
            Network & Accuracy (\%) \\
            \midrule
            COVID-Net~CT-1~\cite{Gunraj2020} & 94.5\\
            COVID-Net~CT-2~L &  \textbf{98.1}\\
            COVID-Net~CT-2~S &  97.9\\
            \bottomrule
        \end{tabular}
        \vspace{-0.1in}
    \end{table}

    \begin{table}[h]
        \caption{Sensitivity for each infection type at the image level on the COVIDx~CT-2 benchmark test dataset.  Best results highlighted in \textbf{bold}.}
        \medskip
        \label{table:sens}
        \centering
        \begin{tabular}{llll} \hline
            \toprule
            Network & \multicolumn{3}{c}{Sensitivity (\%)}\\
            \cmidrule(lr){2-4}
             & Normal & CP & NCP \\
            \cmidrule(lr){2-4}
            COVID-Net~CT-1~\cite{Gunraj2020} & 98.8 & \textbf{99.0} & 80.2 \\
            COVID-Net~CT-2~L & \textbf{99.0} & 98.2 & \textbf{96.2} \\
            COVID-Net~CT-2~S & 98.9 & 98.1 & 95.7\\
            \bottomrule
        \end{tabular}
    \end{table}

    \begin{table}[h]
        \caption{Positive predictive value (PPV) for each infection type at the image level on the COVIDx~CT-2 benchmark test dataset.  Best results highlighted in \textbf{bold}.}
        \medskip
        \label{table:ppv}
        \centering
        \begin{tabular}{llll} \hline
            \toprule
            Network & \multicolumn{3}{c}{PPV (\%)} \\
            \cmidrule(lr){2-4}
             & Normal & CP & NCP \\
            \cmidrule(lr){2-4}
            COVID-Net~CT-1~\cite{Gunraj2020} & 96.1 & 90.2 & \textbf{97.6}\\
            COVID-Net~CT-2~L & \textbf{99.4} & \textbf{97.2} & 96.7\\
            COVID-Net~CT-2~S & 99.3 & 97.0 & 96.4\\
            \bottomrule
        \end{tabular}
    \end{table}

    \begin{table}[h]
        \caption{Specificity for each infection type at the image level on the COVIDx~CT-2 benchmark test dataset.  Best results highlighted in \textbf{bold}.}
        \medskip
        \label{table:spec}
        \centering
        \begin{tabular}{llll} \hline
            \toprule
            Network & \multicolumn{3}{c}{Specificity (\%)} \\
            \cmidrule(lr){2-4}
             & Normal & CP & NCP \\
            \cmidrule(lr){2-4}
            COVID-Net~CT-1~\cite{Gunraj2020} & 96.3 & 95.7 & \textbf{99.4}\\
            COVID-Net~CT-2~L & \textbf{99.5} & \textbf{98.8} & 99.0\\
            COVID-Net~CT-2~S & 99.3 & 98.8 & 98.9\\
            \bottomrule
        \end{tabular}
    \end{table}

    \begin{table}[h]
        \caption{Negative predictive value (NPV) for each infection type at the image level on the COVIDx~CT-2 benchmark test dataset.  Best results highlighted in \textbf{bold}.}
        \medskip
        \label{table:npv}
        \centering
        \begin{tabular}{llll} \hline
            \toprule
            Network & \multicolumn{3}{c}{NPV (\%)} \\
            \cmidrule(lr){2-4}
             & Normal & CP & NCP \\
            \cmidrule(lr){2-4}
            COVID-Net~CT-1~\cite{Gunraj2020} & 98.9 & \textbf{99.6} & 94.2\\
            COVID-Net~CT-2~L & \textbf{99.1} & 99.3 & \textbf{98.8}\\
            COVID-Net~CT-2~S & 99.0 & 99.2 & 98.7\\
            \bottomrule
        \end{tabular}
    \end{table}

\subsection{Qualitative analysis}

\begin{figure}[t]
        \centering
        \includegraphics[width=0.95\textwidth]{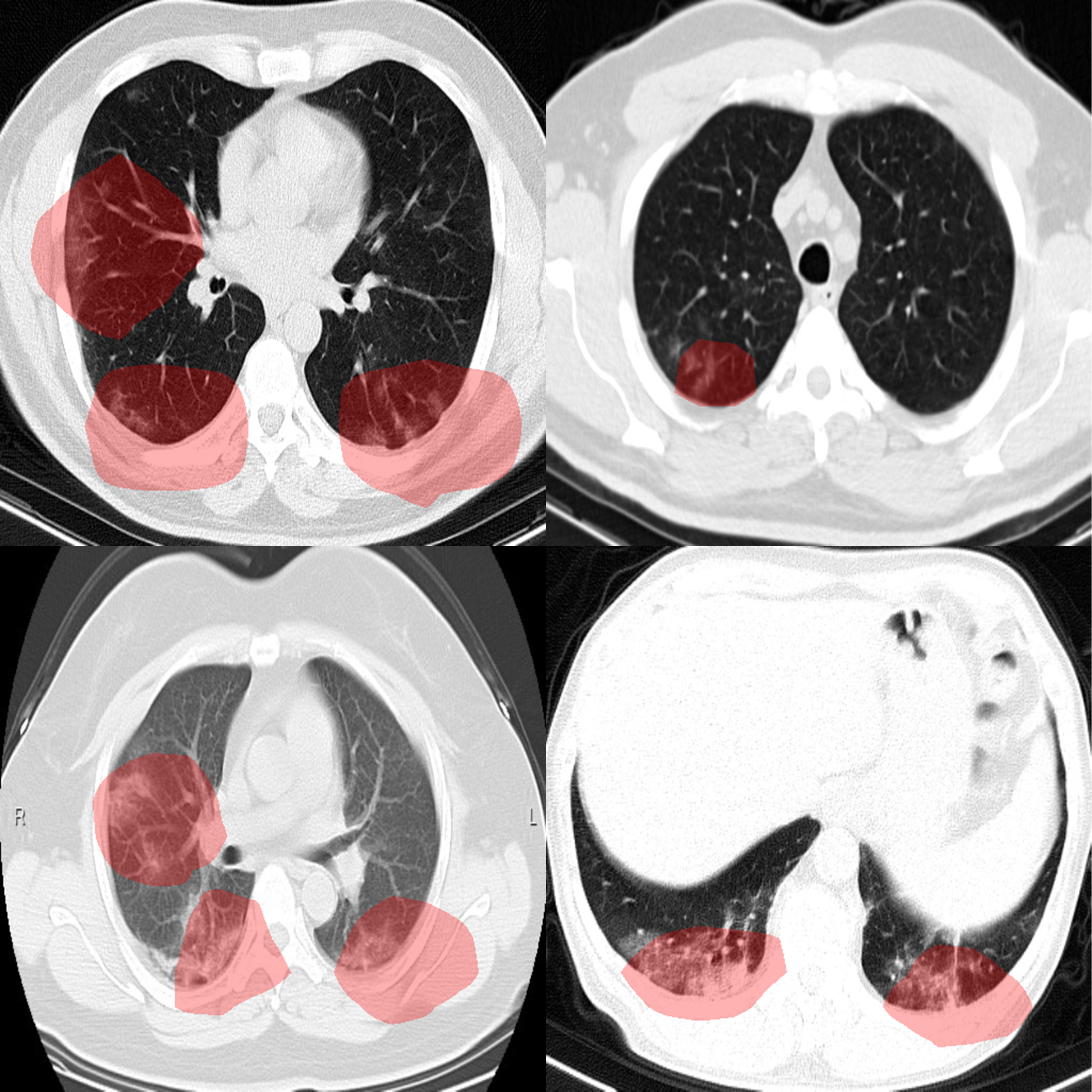}
        \caption{Example chest CT images from four COVID-19 cases, and the associated critical factors (highlighted in red) as identified by GSInquire~\cite{gsinquire} for COVID-Net~CT-2~S.}
        \label{fig:explain2}

    \end{figure}

    To audit the decision-making behaviour of COVID-Net~CT-2 and ensure that it is leveraging relevant visual indicators when predicting the condition of a patient, we conducted explainability-driven performance validation using the COVIDx~CT-2 benchmark test dataset, and the results obtained using COVID-Net~CT-2~L for select patient cases are further reviewed and reported on by two board-certified radiologists. The critical factors identified by GSInquire for example chest CT images from the four COVID-19-positive cases that were reviewed are shown in Figure~\ref{fig:explain}, and additional examples for COVID-Net~CT-2~S are shown in Figure~\ref{fig:explain2}.

    Overall, it can be observed from the GSInquire-generated visual explanations that both COVID-Net~CT-2~L and COVID-Net~CT-2~S are mainly utilizing visible lung abnormalities to distinguish between COVID-19-positive and COVID-19-negative cases.  As such, this auditing process allows us to determine that COVID-Net~CT-2 is indeed leveraging relevant visual indicators in the decision-making process as opposed to irrelevant visual indicators such as imaging artifacts, artificial padding, and patient tables. This performance validation process also reinforces the importance of utilizing explainability methods to confirm proper decision-making behaviour in deep neural networks designed for clinical decision support.

\subsection{Radiologist findings}

\begin{figure}[t]
        \centering
        \includegraphics[width=0.95\textwidth]{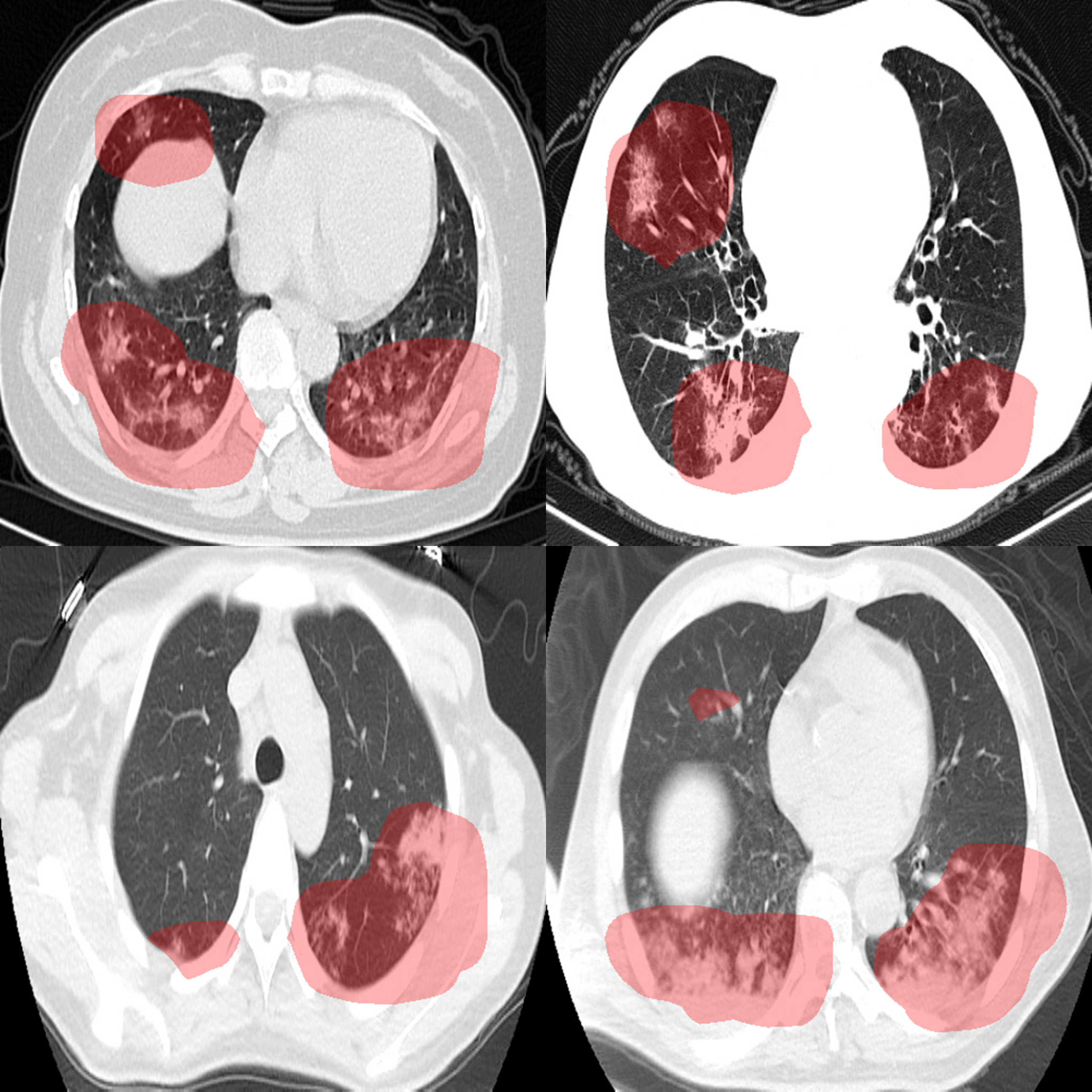}
        \caption{Example chest CT images from four COVID-19 cases reviewed and reported on by two board-certified radiologists, and the associated critical factors (highlighted in red) as identified by GSInquire~\cite{gsinquire} for COVID-Net~CT-2~L.  Based on the observations made by two expert radiologists, it was found that the critical factors leveraged by COVID-Net~CT-2~L are consistent with radiologist interpretation.}
        \label{fig:explain}

    \end{figure}

    The expert radiologist findings and observations with regards to the critical factors identified by GSInquire for each of the four patient cases shown in Figure~\ref{fig:explain} are as follows. In all four cases, COVID-Net~CT-2~L detected them to be novel coronavirus pneumonia due to SARS-CoV-2 viral infection, which was clinically confirmed.

    \textbf{Case 1 (top-left of Figure~\ref{fig:explain})}. It was observed by one of the radiologists that there are bilateral peripheral mixed ground-glass and patchy opacities with subpleural sparing, which is consistent with the identified critical factors leveraged by COVID-Net~CT-2~L.  The absence of large lymph nodes and effusion further helped the radiologist point to novel coronavirus pneumonia due to SARS-CoV-2 viral infection.  The degree of severity is observed to be moderate to high.  It was confirmed by the second radiologist that the identified critical factors leveraged by COVID-Net~CT-2~L are correct areas of concern and represent areas of consolidation with a geographic distribution that is in favour of novel coronavirus pneumonia due to SARS-CoV-2 viral infection.

    \textbf{Case 2 (top-right of Figure~\ref{fig:explain})}. It was observed by one of the radiologists that there are bilateral peripherally-located ground-glass opacities with subpleural sparing, which is consistent with the identified critical factors leveraged by COVID-Net~CT-2~L.  As in Case 1, the absence of large lymph nodes and large effusion further helped the radiologist point to novel coronavirus pneumonia due to SARS-CoV-2 viral infection.  The degree of severity is observed to be moderate to high.  It was confirmed by the second radiologist that the identified critical factors leveraged by COVID-Net~CT-2~L are correct areas of concern and represent areas of consolidation with a geographic distribution that is in favour of novel coronavirus pneumonia due to SARS-CoV-2 viral infection.

    \textbf{Case 3 (bottom-left of Figure~\ref{fig:explain})}. It was observed by one of the radiologists that there are peripheral bilateral patchy opacities, which is consistent with the identified critical factors leveraged by COVID-Net~CT-2~L.  Unlike the first two cases, there is small right effusion.  However, as in Cases 1 and 2, the absence of large effusion further helped the radiologist point to novel coronavirus pneumonia due to SARS-CoV-2 viral infection. Considering that the opacities are at the base, a differential of atelectasis change was also provided. The degree of severity is observed to be moderate.  It was confirmed by the second radiologist that the identified critical factors leveraged by COVID-Net~CT-2~L are correct areas of concern and represent areas of consolidation.

    \textbf{Case 4 (bottom-right of Figure~\ref{fig:explain})}.  It was observed by one of the radiologists that there are peripherally located asymmetrical bilateral patchy opacities, which is consistent with the identified critical factors leveraged by COVID-Net~CT-2~L.  As in Cases 1 and 2, the absence of lymph nodes and large effusion further helped the radiologist point to novel coronavirus pneumonia due to SARS-CoV-2 viral infection, but a differential of bacterial pneumonia was also provided considering the bronchovascular distribution of patchy opacities. In addition, there is no subpleural sparing.  This highlights the potential difficulties in differentiating between novel coronavirus pneumonia and common pneumonia.  It was confirmed by the second of the radiologists that the identified critical factors leveraged by COVID-Net~CT-2~L are correct areas of concern and represent areas of consolidation with a geographic distribution that is in favour of novel coronavirus pneumonia due to SARS-CoV-2 viral infection.

    Therefore, it can be observed that the explainability-driven validation process shows consistency between the decision-making process of COVID-Net CT-2 and radiologist interpretation, which suggests strong potential for computer-aided COVID-19 assessment within a clinical environment.

    Based on both quantitative and qualitative results, it can be seen that not only does COVID-Net~CT achieve high performance, but it is  leveraging relevant abnormalities in the lungs in its decision-making process rather than erroneous visual cues.

\section{Conclusions and Discussion}

    In this work, we introduced COVID-Net~CT-2, enhanced deep convolutional neural networks tailored for the purpose of COVID-19 detection from chest CT images via more diverse learning on the largest quantity and diversity of multinational patient cases in research literature.  Two new CT benchmark datasets were introduced and used to facilitate the learning of COVID-Net~CT-2, and these datasets represent the largest, most diverse, multinational cohorts of their kind available in open access form, spanning cases from at least 15 countries.  Experimental results show that the COVID-Net~CT-2 networks are capable of not only achieving strong test accuracy, sensitivity, and positive predictive value, but also do so in a manner that is consistent with radiologist interpretation via explainability-driven performance validation.  The results are promising and suggest the strong potential of deep neural networks as an effective tool for computer-aided COVID-19 assessment.

    Given the severity of the COVID-19 pandemic and the potential for deep learning as a potential tool to facilitate computer-assisted COVID-19 clinical decision support, a number of deep learning systems have been proposed in research literature for detecting SARS-CoV-2 infections using CT images~\cite{Mei2020, cncb, Xu2020, Bai2020_Aug, Li2020, Ardakani2020, Shah2020, Chen2020, Zheng2020, Jin2020, Jin2020_2, Song2020, Wang2020_Screen,Harmon2020,Gunraj2020,HUST}. While some proposed deep learning systems focus on binary detection (SARS-CoV-2 positive vs. negative)~\cite{Harmon2020}, several proposed systems operate at a finer level of granularity by further identifying whether SARS-CoV-2 negative cases are normal control~\cite{cncb, Xu2020, Jin2020_2, Song2020}, SARS-CoV-2 negative pneumonia (e.g., bacterial pneumonia, viral pneumonia, community-acquired pneumonia (CAP), etc.)~\cite{cncb, Xu2020, Bai2020_Aug, Li2020, Ardakani2020, Song2020, Wang2020_Screen}, or non-pneumonia~\cite{Li2020}.

    The majority of the proposed deep learning systems for COVID-19 detection from CT images rely on pre-existing network architectures that were designed for other image classification tasks. A large number of proposed systems additionally rely on segmentation of the lung region and/or lung lesions~\cite{Mei2020, cncb, Xu2020, Bai2020_Aug, Li2020, Chen2020, Zheng2020, Jin2020_2, Song2020}. Some proposed systems also augment pre-existing network architectures, with Xu et al.~\cite{Xu2020} augmenting a pre-existing ResNet-18~\cite{resnet} backbone architecture with location-attention classification, and Li et al.~\cite{Li2020} and Bai et al.~\cite{Bai2020_Aug} augmenting pre-existing network architectures with pooling operations for volume-driven detection. Of the deep learning systems that proposed new deep neural network architectures, Shah et al.~\cite{Shah2020} proposed a 10-layer convolutional neural network architecture named CTnet-10, which ultimately showed lower detection performance than pre-existing architectures in literature. Zheng et al.~\cite{Zheng2020} proposed a 3D convolutional neural network architecture named DeCovNet which is capable of volume-driven detection. Finally, in the system introduced by Gunraj et al.~\cite{Gunraj2020}, machine-driven design exploration was leveraged to construct a deep neural network architecture that is tailored specifically for COVID-19 detection using CT images.

    While the concept of leveraging deep learning for COVID-19 detection from CT images has been previously explored, even the largest studies in research literature in this area have been limited in terms of quantity and/or diversity of patients, with many limited to single-nation cohorts. For example, the studies by Mei et al.~\cite{Mei2020}, Gunraj et al.~\cite{Gunraj2020}, Ning et al.~\cite{HUST}, and Zhang et al.~\cite{cncb} were all limited to Chinese patient cohorts consisting of 905 patients, 1,489 patients, 1,521 patients, and 3,777 patients, respectively. The largest multinational study in research literature was conducted by Harmon et al.~\cite{Harmon2020}, which leveraged a cohort of 2,617 patients across 4 countries. To the best of the authors' knowledge, the largest of the unified multinational patient cohorts introduced in this study represents the largest, most diverse multinational patient cohort at 4,501 patients across at least 15 countries. By building the proposed COVID-Net CT-2 deep neural networks using a large multinational patient cohort, we can better study the generalization capabilities and applicability of deep learning for computer-assisted assessment under a wider diversity of clinical scenarios and demographics.

    With the tremendous burden the ongoing COVID-19 pandemic has put on healthcare systems and healthcare workers around the world, the hope is that research such as COVID-Net~CT-2 and open source initiatives such as the COVID-Net initiative can accelerate the advancement and adoption of deep learning solutions within a clinical setting to aid front-line health workers and healthcare systems in improving clinical workflow efficiency and effectiveness in the fight against the COVID-19 pandemic.  While to the best of the authors' knowledge this research does not put anyone at a disadvantage, it is important to note that COVID-Net~CT-2 is not a production-ready solution and is meant for research purposes.  As such, predictions made by COVID-Net~CT-2 should not be utilized blindly and should instead be built upon and leveraged in a human-in-the-loop fashion by researchers, clinicians, and citizen data scientists alike.  Future work involves leveraging the core COVID-Net~CT-2 backbone for downstream tasks such as lung function prediction, severity assessment, and actionable predictions for guiding personalized treatment and care for SARS-CoV-2 positive patients.

\begin{ack}
    We thank the Natural Sciences and Engineering Research Council of Canada (NSERC), the Canada Research Chairs program, the Canadian Institute for Advanced Research (CIFAR), DarwinAI Corp., Justin Kirby of the Frederick National Laboratory for Cancer Research, and the various organizations and initiatives from around the world collecting valuable COVID-19 data to advance science and knowledge. The study has received ethics clearance from the University of Waterloo (42235).
\end{ack}

\section*{Author contributions statement}
H.G. and A.W. conceived the experiments, H.G. conducted the experiments, all authors analysed the results, D.K. and A.S. reviewed and reported on select patient cases and corresponding explainability results illustrating model's decision-making behaviour, and all authors reviewed the manuscript.

\section*{Declaration of interests}
A.W. is affiliated with DarwinAI Corp.

\bibliographystyle{IEEEtran}
\bibliography{references}

\end{document}